\documentclass[doublecol,showpacs,preprintnumbers,amsmath,cite,linenumbers]{epl2}
\usepackage[normalem]{ulem}
\usepackage{graphicx}% Include figure files
\usepackage{dcolumn}% Align table columns on decimal point
\usepackage{bm}% bold math
\usepackage{amsmath}
\usepackage{amssymb}
\usepackage{color}
\usepackage[switch]{lineno}
\begin{document}

\title{Orientation, Flow, and Clogging  in a Two-Dimensional Hopper:
  Ellipses vs. Disks}

\author{Junyao Tang\inst{1,2} \and
R. P. Behringer\inst{1}}

\shortauthor{J. Tang and R. Behringer}

\institute{
\inst{1} Department of Physics \& Center for Nonlinear and Complex
Systems, Duke University, Durham, NC, 27708, USA\\

\inst{2} 10451 Clay Road, TGS-Nopec Geophysical Co., Houston, TX, 77043, USA\\

}

\abstract{Two-dimensional (2D) hopper flow of disks has been
  extensively studied.  Here, we investigate hopper flow of ellipses
  with aspect ratio $\alpha = 2$, and we contrast that behavior to the
  flow of disks.  We use a quasi-2D hopper containing photoelastic
  particles to obtain stress/force information. We simultaneously
  measure the particle motion and stress. We determine several
  properties, including discharge rates, jamming probabilities, and
  the number of particles in clogging arches. For both particle types,
  the size of the opening, $D$, relative to the size of particles,
  $\ell$ is an important dimensionless measure. The orientation of the
  ellipses plays an important role in flow rheology and clogging.  The
  alignment of contacting ellipses enhances the probability of forming
  stable arches. This study offers insight for applications involving
  the flow of granular materials consisting of ellipsoidal shapes, and
  possibly other non-spherical shapes.}

\date{\today}

\pacs{47.57.Gc}{Granular flow}
\pacs{81.05.Rm}{Porous materials; granular materials}
\pacs{78.20.Fm}{Birefringence}

\newcommand{\RNum}[1]{\uppercase\expandafter{\romannumeral #1\relax}}
\maketitle

% MAIN TEXT 

%\linenumbers 

Hopper flows of granular materials involve dynamical granular states
with important industrial applications\cite{Nedderman,
  industrial}. Time-averaged granular flow theories, often using
hopper flow as a test case, have progressed from continuum mechanics
models to mesoscopic models ~\cite{Nedderman, RaoNott, Bazant, Kamrin,
  GDRMidi}.  Fluctuations and clogging (or jamming) are also important
characteristics for hopper flow. Experiments~\cite{Zuriguel2, Durian,
  Durian2} have examined the the clogging transition of hopper flow
for different grain properties and hopper geometries. Most
recent results from~\cite{Durian} imply that all hoppers have a
nonzero probability to clog. Other studies~\cite{Menon2, Vivanco, To1,
  To2, Ferguson_PRE06, ifpri} have also sought to understand flow and
clogging (or jamming) mechanisms from a microscopic viewpoint.
However, for simplicity, theories developed from the above studies
often tend to assume spherical particles, including disks in two
dimensions (2D). The effect of particle shape on hopper flow is
usually not their focus. In reality, particle shapes are often not
spherical; rice and M$\&$M's are roughly ellipsoids; sand particles
have irregular shapes. Thus, it is scientifically and industrially
relevant to explore how particle shape affects flow rheology and
clogging mechanisms.

A simple way to explore particle shape effects is to contrast 2D
hopper flows of disks and ellipses for the time-averaged discharge
rate, $\dot{M}$, and jamming probability. Discharge rates of hopper
flow often follow the well-established Beverloo
equation\cite{Beverloo_CES61}, which relates $\dot{M}$ to the hopper
opening size, $D$~:~$\dot{M} \propto (D - kd_{avg})^{(n -1/2)}$, where
$n$ is the spatial dimension (e.g. $n = 2 \text{ or } 3$ for two or
three-dimensional systems). $D$ is reduced by $kd_{avg}$ due to
boundary effects, where $d_{avg}$ characterizes the grain size, and
$k$ is an order-one constant~\cite{Nedderman}. Recent studies
\cite{JandaMaza:2012, JandaMaza:2015} have provided micromechanical
insights into this equation with a coarse-grain technique. An
important open question concerns the relevance of this relation for
non-spherical particles. Several studies have used DEM simulation
methods to understand how the aspect ratio of an ellipse could affect
the discharge rate~\cite{CLEARY:1999, Tao:2011, CLEARY:2002,
  Langston:2004, Li:2004, Campbell:2009, Hohner:2013}. Their results
are not consistent due to different assumptions such as particle
shapes, frictional properties of particles and etc.  A recent study by
Liu et al.~\cite{Liu:2014} utilizing both DEM and experiments,
suggests a modified Beverloo equation for ellipse flow. They also show
observations of flow characteristics based on their simulation
results. To our knowledge, experimental investigations of hopper flow
of ellipses based on particle-levels dynamics are still lacking.

%Briefly, the Beverloo equation assumes that the material discharge speed, $V$, at the outlet follows $V \propto [g(D - kd_{avg})]^{1/2}$.  $\dot{M}$ is then

%The Beverloo equation works well for hopper flows of dry granular materials that are roughly spherical\cite{Nedderman, RaoNott}.  For spherical particles, less is known about clogging probabilities, and the dependence of $\tau_c$ on $D$ is still the subject of ongoing study\cite{Durian}. Much less is known about how particle shape affects flow rates and clogging probabilities. 

In this paper we experimentally test the applicability of the Beverloo equation to
elliptical particles in quasi-2D hopper flows. We also measure the probablity of jamming for these particles. We
contrast these quantities to results for bi-disperse
disks in the same hopper. To apply the Beverloo equation, an
issue arises for elongated particles such as ellipses: since there
are two lengths for the particles, the
major and minor principal axis lengths, $d_{maj}$ and $d_{min}$
respectively, which if either is relevant? Thus, there are two dimensionless length
ratios, which can be taken as $D/d_{min}$ and the ellipse aspect
ratio, $\alpha = d_{maj}/d_{min}$. In addition, the organization of
elliptical grains, and the stable structures that they form when a
jammed state occurs following a clog are important.

In the present experiments, we use a 2D wedge-shaped hopper, as in our
previous studies ~\cite{Tang:2009, ifpri}. The apparatus consists of
two transparent Plexiglas sheets separated by aluminum spacers and
aluminum hopper walls. The
system is divided into upper and lower regions, both of which are
hoppers. Approximately 5,000 bi-disperse circular particles or 3,000
identical elliptical particles are initially placed in the upper
hopper section. For ellipses, $d_{maj}=$~10~mm, and $d_{min}
=$~5~mm. For disks, we have big particles (diameter $d =$~6~mm ) and
small particles (diameter $d =$~5~mm ) with relative fraction small to
large of 2:1. So the average diameter of the bi-disperse circular
particles (disks) is 5.3 mm.  Two sliding Teflon bars initiate or stop
the flow. The upper hopper can be reloaded easily by rotating twice
about a pivot. Here, we consider results for a hopper wall angle
$\theta_w$=$30^o$ ($\theta_w$ is half the full opening angle of the
hopper).  The particles are made of photoelastic materials (Vishay,
PSM); when they are placed between crossed polarizers, transmitted
light produces images with fringe-like patterns that depend on the
forces acting on each particle~\cite{Tang:2011}. We use the
photoelastic response to measure grain-scale forces. The square of the
image intensity gradient provides a measure of the local force, which
we calibrate by applying known static loads to systems of
particles~\cite{Howell:1999}. Note that the contact forces between
particles can also be accurately calculated based on photoelastic
principles, for high-resolution images ~\cite{Trush:2005}. This is not
possible in the present experiments that rely on high speed images
with modest frame sizes.

\begin{figure}
\centering
            \includegraphics[width=6.0cm]{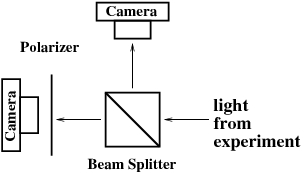}
            \caption{\label{Fig: Experiment} Sketch of the
              synchronized camera set up.  Light emerging from the
              hopper is split and imaged by two synchronized
              high-speed cameras, one with and the other without a
              crossed polarizer.}
\end{figure}

\begin{figure}
         \includegraphics[width=8cm,height=17.5cm]{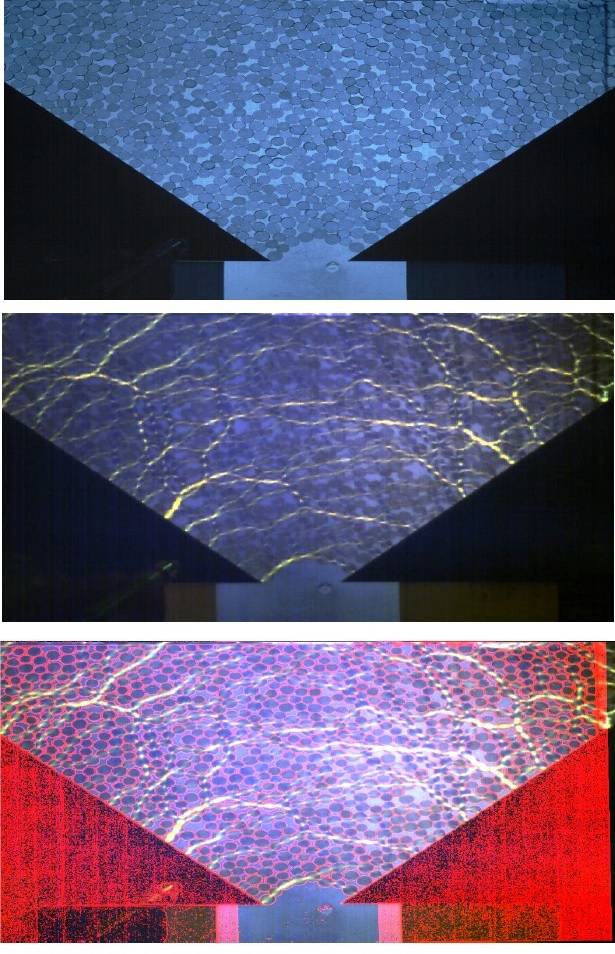}
           \caption{\label{Fig: ExperimentPTGSQ} (Color
             Online). Sample images from the top cameras, and their
             overlap. Top: A direct image. Middle: The corresponding
             polarized image. Bottom: The image that results from
             overlapping the top and middle images. }
\end{figure}

In order to simultaneously observe particles positions and the forces
acting on them, we use two synchronized high-speed cameras (frame
rate: 500fps), one to take pictures for particle tracking
(Fig.~\ref{Fig: ExperimentPTGSQ}-top) and the other for photoelastic
measurements (Fig.~\ref{Fig: ExperimentPTGSQ}-middle). As in
Fig.~\ref{Fig: Experiment}, a beam splitter steers polarized light
from the experiment into the two cameras.  One camera has a polarizer
that is crossed with respect to the original light polarization. The
other lacks a second polarizer, and only registers the direct images
of the particles. The images from the cameras are aligned through
registration techniques to produce a composite image that details the
location and orientation of particles, and the photoelastic response.
Note that photoelastic images alone cannot be used to locate
particles, since particles and/or their boundaries are often
invisible. Fig.~\ref{Fig: ExperimentPTGSQ} shows two original (direct
and polarized) images from the two cameras, and the resulting
overlapped image. This figure shows the force chains corresponding to
strongly stressed particles. Below, we use this dual information to
understand key differences between the flow and clogging of ellipses
vs. disks.

\begin{figure}
            \includegraphics[width=8cm,height=5.5cm]{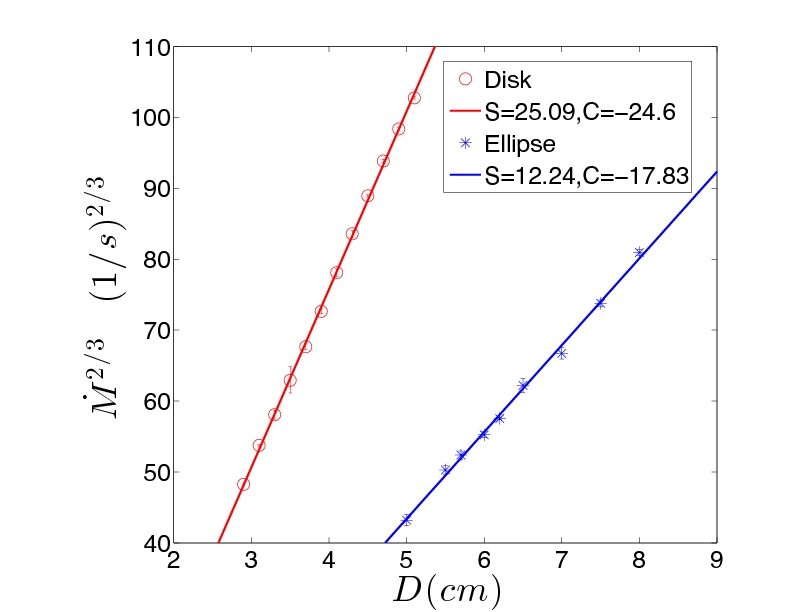}
            \caption{\label{Fig: ElpDischargeRate} (Color Online). Discharge rate vs  opening sizes of disks and ellipses. (Hopper wall angle $\theta_w={30}^o$). }%Inset: Discharge rate vs  normalized opening sizes of disks and ellipses. For disks, $d=d_c=0.53cm$; for ellipses1, $d=d_{maj}=1cm$.}
\end{figure}

\begin{figure}
            \includegraphics[width=8cm,height=5.5cm]{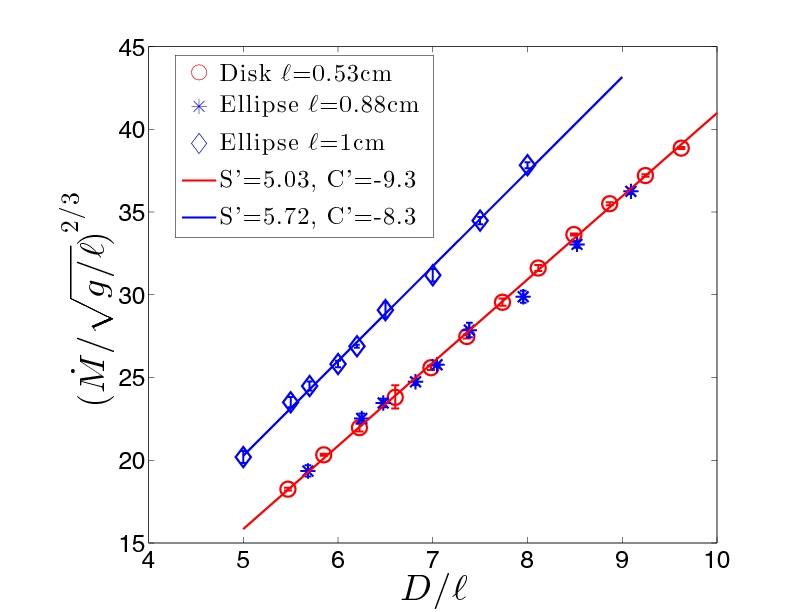}
            \caption{\label{fig:m-dot-dimensionless} (Color
              Online). Dimensionless discharge rate vs. dimensionless
              opening size for disks and ellipses. (Hopper wall angle
              $\theta_w={30}^o$). }%Inset: Discharge rate vs
            %normalized opening sizes of disks and ellipses. For disks,
            %$d=d_c=0.53cm$; for ellipses1, $d=d_{maj}=1cm$.}
\end{figure}

We start by comparing the time-averaged discharge rate for disks and
ellipses. We measured $\dot{M}$ as the ratio of the total number of particles
to the total time taken to empty the hopper. If a jam occurred, we
re-initiated the flow by controlled taps with a small hammer located
outside the Plexiglas. The total time to empty the material is the sum
of the consecutive times during which the particles flowed. In
Fig.~\ref{Fig: ElpDischargeRate}, we show the discharge rate raised to
the $2/3$ power, $\dot{M}^{2/3}$, vs. the opening size, $D$. If the
Beverloo equation holds, there should be a linear relation (in 2D) for
$\dot{M}^{2/3}$ vs. $D$~\cite{Beverloo_CES61}. The data for both disks
and ellipses are consistent with such a relation, $\dot{M}^{2/3} = S D
+ C$, and hence the Beverloo equation is satisfied (The fitting
parameters S and C appear in Fig.~\ref{Fig:
  ElpDischargeRate}~and Fig.~\ref{fig:m-dot-dimensionless}).  In physical
units, the ellipses flow more slowly than the disks at the same
opening size; $\dot{M}$ data in Fig.~\ref{Fig: ElpDischargeRate} for
the ellipses have a slope that is close to 1/2 that for the
disks. Where might such a difference arise?  Also, what is a
reasonable way to compare these two data sets, given that the minor
axis of the ellipses is comparable to the diameter of the disks, but
the major axis is roughly twice the disk diameter? The Beverloo
equation does not provide insight (except in a rough way through the
boundary layer term, $gkd_{avg}$ ) into the role of particle
shape.

One approach is to seek non-dimensional rescaled representations of
the data for $\dot{M}$ and $D$. The former, has dimensions of inverse
time. The time scale must come from $g$ and a length scale related to
the particle size.  Similarly, the scale for $D$ involves a
particle-scale length. For simplicity, we assume that the same
measure, $\ell_i$, which depends on the particle species ($i = d$ for
disks or $i = e$ for ellipses) applies for both $\dot{M}$ and $D$. We
write $\dot{M} = (g/\ell_i)^{1/2} \dot{M}'$ and $D = \ell_i D'$, where
the primed quantities are dimensionless. Since $\dot{M}^{2/3} \propto
D$, for dimensionless quantities: $\dot{M}'^{2/3} = (\ell_i/g)^{1/3}\dot{M}^{2/3} \propto \ell_i^{1/3} 
\ell_i  D'$. There is a universal expression
$\dot{M}'^{2/3}=S'D'+C'$ if $\ell_e^{4/3} S_e /\ell_d^{4/3} S_d =
1$. The measured slopes for ellipses and disks satisfy $S_d/S_e \simeq
2$, which yields $\ell_e^{4/3}/\ell_d^{4/3} \simeq 2$ This is roughly
satisfied if $\ell_d = d$ and $\ell_e \simeq d_{maj}$. Then,
$\ell_e^{4/3} /\ell_d^{4/3} = (d_{maj}/d)^{4/3} = 2.3$. This is
reasonably consistent with the ratio of the slopes in Fig.~\ref{Fig:
  ElpDischargeRate}, $S_d/S_e = 2.0 \pm 0.2$.

Fig.~\ref{fig:m-dot-dimensionless} shows data for the dimensionless
flow rate $\dot{M}'$ vs. dimensionless hopper opening, $D'$, for $\ell_e
= 1.0$~cm~$= d_{maj}$ as the length scale factor, and a best fit
$\ell_e = 0.88$~cm.  For the latter choice of $\ell_e$, the collapse of
the disk and ellipse flow rate data is complete, within the scatter.

\begin{figure}[!ht]
            {\includegraphics[width=8cm,height=5.5cm]{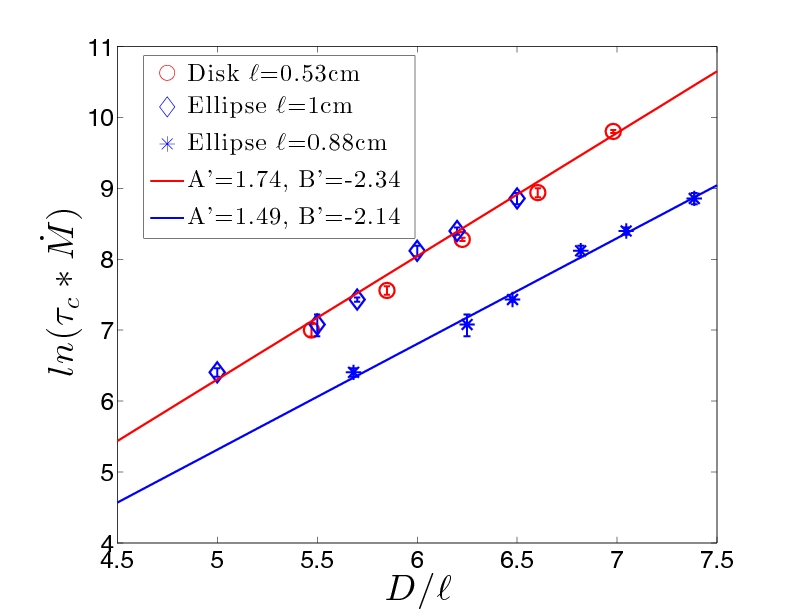}}
            \caption{\label{fig:tau-dimsionless} (Color
              Online). Semi-log plot of ${\tau}_c \dot{M}$ vs opening
              size for both disks and ellipses (hopper wall angle
              $\theta_w={30}^o$). Dimensionless version. Stars correspond to the dimensionless data with the length scale $\ell_e = 0.88$~cm that collapse disks and ellipses discharge rate data in Fig 3.}
\end{figure}

Previously~\cite{Tang:2009}, we showed for our disks that
flowing/clogging can be described as a Poisson process. If the
probability of flow without a clog in time $dt$ is $dt/\tau_c$, then
the probability that the flow persists without clogging until time $t$
is $P = \exp(-t/\tau_c)$, where the survival time $\tau_c$ reflects
the the jamming probability of hopper flow.  In Fig.~\ref{Fig:
  ElpJam}, we compare the jamming probability of a system of disks and
a system of ellipses, where, we use ${\tau}_c \dot{M}$ (at a given
$D$), i.e., the average number of particles that fall out before
jamming, similar to the term  ``$n$" used in ~\cite{To2}. For both particle types,
${\tau}_c \dot{M}$ grows strongly with $D$, consistent with
exponential dependence, i.e., $ln({\tau}_c \dot{M}) =A D +B$. Although
the experimentally accessible ranges of $D$ for the two data types do
not overlap, extrapolation suggests that ellipse flows jam more
readily than disk flows at the same $D$. Like the flow rate, it is
interesting to rescale the physical quantities in Fig.~\ref{Fig:
  ElpJam}. The vertical axis is already dimensionless. If we rescale
$D$ by the mean diameter of the disks and by the major diameter of the
ellipses, we obtain a good collapse of the disk and ellipse data for
${\tau}_c \dot{M}$, as shown in Fig.~\ref{fig:tau-dimsionless}
($ln({\tau}_c \dot{M}) =A' D' +B'$), where the primes here refer to
fits to the scaled dimensionless data.

%%%\begin{figure}[!htt]
        %%\begin{center}
          %%%  \includegraphics[width = 7.5cm,height=5cm]{PdfTau_Elp.jpg}
        %%%    \caption{\label{Fig: ElpJamExp} Probability distribution of survival time of D=5cm, ${\theta}_w=60^{o}$ for ellipses, fitted by an exponential function.}      
        %%\end{center}
%%%\end{figure}

\begin{figure}[!ht]
            {\includegraphics[width=8cm,height=5.5cm]{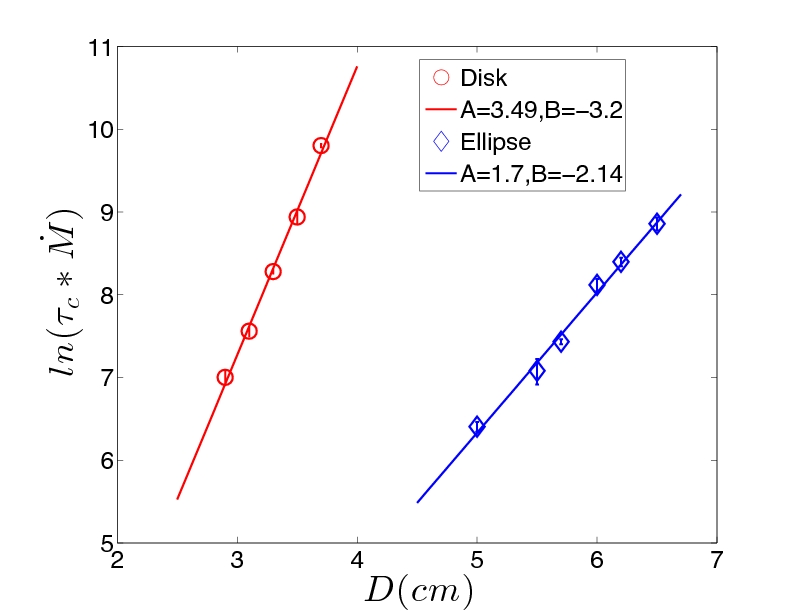}}
            \caption{\label{Fig: ElpJam} (Color Online). Semi-log plot of  ${\tau}_c \dot{M}$  vs  opening size for both disks and ellipses (hopper wall angle $\theta_w={30}^o$). }
\end{figure}

As discussed previously~\cite{Tang:2011, Menon2,
   Ferguson_PRE06}, the formation of transient force
chains during hopper flow is related to the stick-slip events of
hopper flow, which in turn control the flow rate and rheology. Hence,
we expect that a larger probability of forming long-lived force chains
(e.g. near the opening) will be correlated with a lower discharge
rate. 

In the random-walk model of To et al.~\cite{To1, To2}, the probability
of stable blocking arches near the opening depends largely on the
number of particles in the arch. If we assume that $D$ relative to
particle size is the only relevant factor, we might argue that for the
same hopper opening size, the hopper flow of ellipses needs fewer
particles to form the blocking arch than the hopper flow of disks,
since for $\dot{M}$, the microscopic length scale for the ellipses is
double that for disks.  Alternatively, if a flow of ellipses where the
opening size $D_1$ has the same clogging probability as a flow of
disks where the opening is $D_2$, one might expect that there should
be similar number of particles in the blocking arches for ellipses than for
disks. However, this is not the case. Fig.~\ref{Fig:
  HistNumofParticle} shows statistics for the number of particles in
blocking arches for disks and ellipses, where the $D$'s were chosen
(differently) so that the two systems have similar jamming
probabilities (${\tau}_c \dot{M}$=1097 for disks and 1186 for ellipses,
D=2.9cm for disks and 5.5cm for ellipses). Fig.~\ref{Fig:
  HistNumofParticle} shows that for the ellipses, the number of
particles forming the blocking arch has a wider distribution and
sometimes can be as large as 18 particles. On average, the blocking
arch consists of more particles for ellipses than for disks: 12
ellipses and 8 disks. Hence particle size is not the only
differentiating factor between ellipse flow and disk flow.

\begin{figure}[!ht]
            {\includegraphics[width=7.9cm,height=6.5cm]{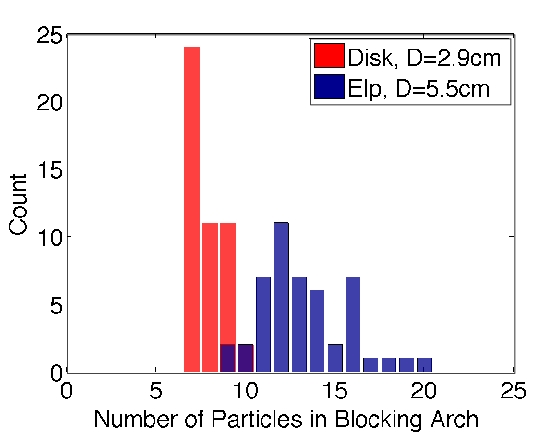}}
            \caption{\label{Fig: HistNumofParticle} (Color Online). Histogram of number of particles in the blocking arch for disks and ellipses.}
\end{figure}

Where does the additional effective stability for ellipses
compared to disks come from?  We address this question using synchronized particle
tracking and photoelastic stress measurements.

\begin{figure}
\centering
            \includegraphics[width=7cm,height=8cm]{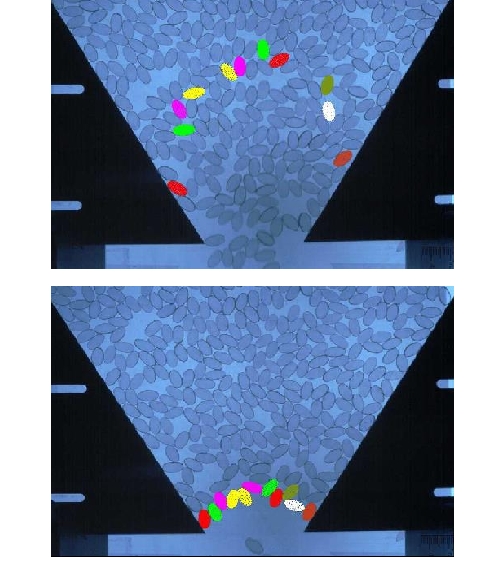}
            \caption{\label{Fig: Elparchtrack} (Color
              Online). Particle-tracking image analysis of sequence of
              a high speed video images showing how apparently random
              particles (brightly colored particles) form the arch
              that stops the flow.  Note that a majority of the
              contacting neighbors for the blocking arch of ellipses tend to
              align roughly parallel.  }
\end{figure}

The source of the difference lies in the fact that ellipses have a coupling between their orientation that affects their
mechanical stability and local density.  Successive frames from the synchronized videos of ellipses,
e.g. Fig.~\ref{Fig: Elparchtrack}, (a complete video is available at:~\cite{elpvideo}), show that ellipse rotation during the flow affects the force chain structure and stability. This rotation is not simply random.

\begin{figure}
\centering
                   {\includegraphics[width=4cm,height=3cm]{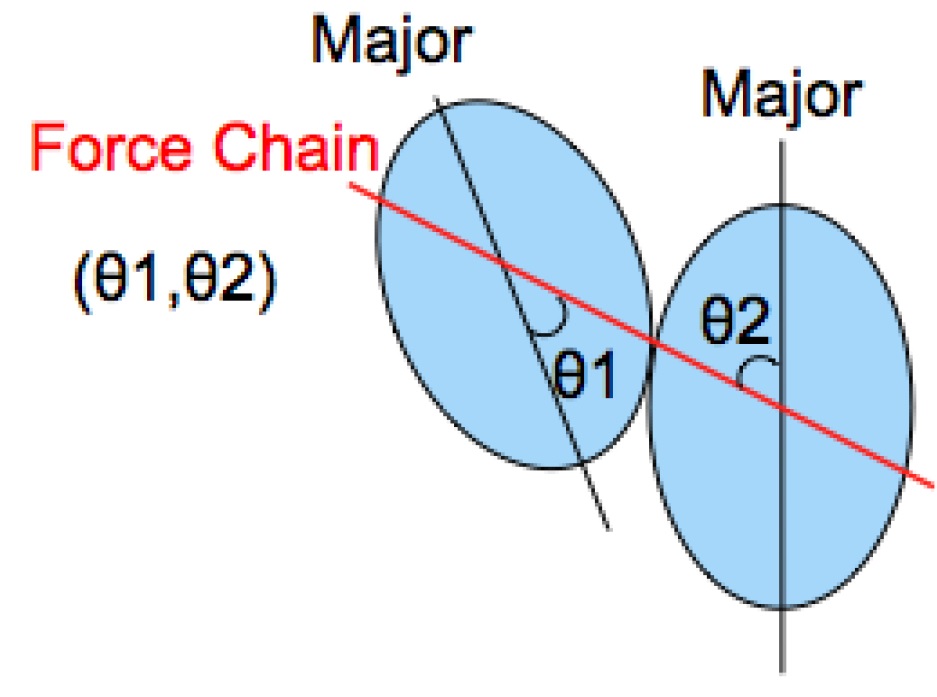}}            
        \caption{\label{Fig: ElpForceChainAnglePair} (Color
          Online). Illustration of the angles
          $({\theta}_1,{\theta}_2)$ that characterize the orientations of contacting ellipses. }%(b) Some examples. Note
        %  $({\theta}_1,{\theta}_2)$ and $({\theta}_2,{\theta}_1)$ are
         % considered as the same contact.}
       
\end{figure}

\begin{figure}[!ht]   
\centering
            {\includegraphics[width=7cm,height=6cm]{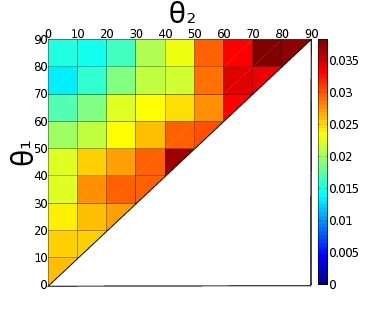}}
             \caption{\label{Fig: ElpAnglePairflow} (Color Online).
               2D histogram of the probability distribution of
               $({\theta}_1,{\theta}_2)$ for stressed ellipses during flow. }
\end{figure}

Specifically, we find a systematic correlation between particle
orientation and force chain orientation, defined below. Force chains
tend to lie along lines corresponding to the local major principal
stress. Using image registration and photoelastic techniques, we
determine this direction by connecting lines through the centers of
the ellipses that experience strong forces.  We determine the mean
force acting on a particle from the gradient-squared measure discussed
above and in previous papers\cite{Howell:1999, Ren:2014}. We
characterize the orientations of contacting ellipses relative to their
contact line, using the angles $({\theta}_1,{\theta}_2)$ illustrated
in Fig.~\ref{Fig: ElpForceChainAnglePair}. Since
  $({\theta}_1,{\theta}_2)$ and $({\theta}_2,{\theta}_1)$ correspond
  to similar cases, we define ${\theta}_1 < {\theta}_2$. We collect
  all angle pairs for force chain particles (defined to be at or above
  the mean force) from approximately 1000 image sequences during a
  flow, and plot the resulting probability distribution in
  Fig.~\ref{Fig: ElpAnglePairflow}. There is a clear orientation
  preference: ellipses forming force chains tend to align parallel to
  their neighbors, with their directions normal to the contact line
  (i.e. the local direction of the force chain).

This parallel preference is even stronger if we limit the analysis to
the force chains that jam the hopper at the outlet. These chains are
usually among the strongest. They form a much smaller data set, but we
show results below for about 100 cases.  Fig.~\ref{Fig:
  ElpAnglePairJam}(a) shows the corresponding
$({\theta}_1,{\theta}_2)$ probability distribution, and Fig.~\ref{Fig:
  ElpAnglePairJam}(b) shows the distribution along the line $\theta_1
= \theta_2$. There is clearly a strong preference for parallel
alignment for neighboring particles in the jamming force chains.

This preferred orientation in strong force chains is a natural
consequence of stability. Two neighboring particles differing
significantly from this orientation, will typically rotate to a more
stable, denser configuration where they are more
parallel\cite{somayeh15}. Simple stability analysis shows that
approximate lines of particles with a parallel configuration can be
stable, even without surrounding particles. These calculations are
supported by a simple test: when multiple ellipses (we have tried up
to 10 particles) are placed on a smooth surface in a parallel
configuration, it is possible to compress the line of particles
without buckling. Such configurations are hypostatic and a similar
test with disks shows that even a very small number of particles in a
line will buckle under uni-axial compression.  This suggests a
starting point for a more detailed quantitative approach to describing
the relation between force networks and ellipse orientation. More
experimental data, such as photoelastic measurements of ellipses with
other aspect ratios, will be helpful for building theoretical models.

\begin{figure}[!ht]
        %\begin{center}
           % \subfigure[]
            %{\includegraphics[width=7cm,height=5cm]{figureThesis/ElpO55JamRun2Compression_6.jpg}}\\
            %\subfigure[]
            %{\includegraphics[width=9cm,height=7cm]{figureThesis/Histogram2DJamElpAngleBetweenElpMaj_ForceChain_Symmetry.jpg}}\\
            %\subfigure[]
            {\includegraphics[width=7.5cm,height=11cm]{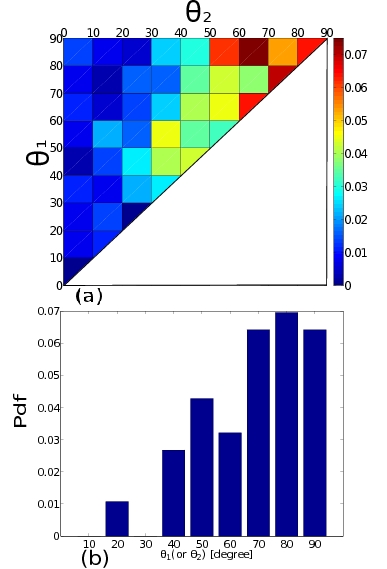}}            
            \caption{\label{Fig: ElpAnglePairJam} (Color Online). (a):
              2D histogram of the probability distribution of
              $({\theta}_1,{\theta}_2)$ for the jammed arch. (b):
              Quantitative plot of the histogram along the diagonal
              line of (a).}
       % \end{center}
\end{figure}

In this paper, we investigated the effect of particle shape on hopper
flow by comparing flow properties of ellipses to those of disks. By
comparing the discharge rate and jamming probability of ellipses to
disks, we find that simple scaling laws allow us to map the flow rates
and jamming probabilities of our elliptical particles onto those for
disks. For both of these properties, the relevant particle length
scale is close to the major diameter of the ellipses. Analysis of the
synchronized particle-tracking and stress data shows that the strongly
stressed elliptical particles that form the strong force chains, tend
to align parallel to their neighbors and transverse to the direction
of the force chains. This effect produces more stable force chains.

Acknowledgements: This work was supported by IFPRI and by NSF grant
DMR-1206351.


\end{document}